\documentstyle[aps]{revtex}

\input{epsf.sty}

\def\vv{{\bf v}}
\def\vr{{\bf r}}
\def\vl{{\bf l}}
\def\vR{{\bf R}}
\def\psib{{\bar \psi}}
\def\vc{{v_{\rm c}}}
\def\vmax{{v_{\rm max}}}
\def\be{\begin{equation}}
\def\ee{\end{equation}}
\def\chie{\chi_{\rm e}}
\begin{document}

\draft

\title{Depinning of a Superfluid Vortex Inside a Circular Defect}

\author{Jos\'e A. Freire}

\address{Departamento de F\'\i sica, Universidade Federal do Paran\'a, \\
Caixa Postal 19081, Curitiba-PR 81531-990, BRAZIL}

\date{\today}

\maketitle

\begin{abstract}

In this work we study the process of depinning of a quantum of circulation 
trapped inside a disk by an applied two dimensional superflow. 
We use the Gross-Pitaevskii model to describe the neutral
superfluid. 
The collective coordinate dynamics is derived directly from the condensate
equation of motion, the nonlinear Schr\"odinger equation, 
and it is used to obtain an expression for the
critical velocity as a function of the defect radius. This expression is
compared with a numerical result obtained from the time independent 
nonlinear Schr\"odinger equation. 
Below the critical velocity, we obtain the dependence of the semiclassical 
nucleation rate with the flow velocity at infinity. 
Above the critical velocity, the classical vortex depinning is
illustrated with a numerical simulation of the time dependent nonlinear 
Schr\"odinger equation.

\end{abstract}


\section{Introduction}

A superflow passing through a constriction or an obstacle of some
sort can produce vorticity in a process known as inhomogeneous nucleation
\cite{Donelly,Hall}. 
The vortex is created at the boundary between the superfluid and the
container walls and two flow regimes are identified. 
When the flow velocity, far away from the walls, is smaller than a
critical velocity an effective barrier is present that prevents vorticity
to appear in the flow, this is the laminar flow regime. 
Above the critical velocity, a vortexless state can dynamically produce
vorticity. In the laminar flow regime, and in the absence of 
thermal excitations,
it is believed that vorticity can still be produced via Macroscopic 
Quantum Tunneling (MQT).

Measurements of the vortex production rates for 
a sphere moving in a superfluid \cite{Hendry} and for an oscillating 
superflow through
a microscopic orifice \cite{Davis,Ihas} have shown evidence of MQT
of vorticity at sufficiently low temperatures. Numerical studies of the
dynamical nucleation process, above the critical velocity, have relied on 
the nonlinear Schr\"odinger model for the superfluid dynamics 
\cite{Frisch,Stone}. 
Analytical studies of the tunneling process have used the collective 
coordinate approach to map what is really a field theory tunneling problem 
into a standard Quantum Mechanical problem for the vortex coordinates 
\cite{Volovik}.
Besides the bosonic superfluid, similar ideas have
been applied to vortex states in short coherence length superconductors 
\cite{Stephen,Gorokhov,Feig,Sonin}.

Different from Ref. \cite{Frisch}, that studied the classical nucleation of a 
vortex/antivortex
pair in the proximity of a circular obstacle in a two dimensional superflow,
in this work we will investigate the depinning of a quantum of circulation
trapped inside the obstacle. 

In the following sections we define
the Gross-Pitaevskii model and extract the dynamics of the collective
coordinate from the condensate equations of motion. We obtain
an expression for the critical velocity and compare it 
with a numerical result. The process of
dynamical nucleation, above the critical velocity, is illustrated with
a numerical integration of the nonlinear Schr\"odinger equation.
The Macroscopic Quantum Tunneling, below the critical velocity, is 
studied within the collective coordinate approach and the semiclassical 
nucleation rate is obtained. 

\section{The Model}

The problem we consider is that of a two dimensional superflow past an 
impenetrable disk perpendicular to the $z$ axis. We will be interested in the 
process of removal of a preexistent quantum of vorticity inside the disk
by the applied superflow (along the $y$ 
direction). In three dimensions this would correspond to a straight vortex
depinning from a columnar defect due to an applied superflow perpendicular
to the column axis. In Ref. \cite{Sonin} this problem is studied 
considering that the vortex depins in the form of an arc attached to the
column. 

We start from the Gross-Pitaevskii model
for the dynamics of the condensate wave function, {\it i.e.} the
nonlinear Schr\"odinger equation (NLSE) \cite{Gross,Pitaevskii}:

\be\label{NLSE}
i\hbar \frac{\partial \psi}{\partial t} = - \frac{\hbar^2}{2m} \nabla^2 \psi
+ g (|\psi|^2 - \rho_0)\psi .
\ee

Here $m$ is the boson mass, $g$ is the parameter in the hard core
potential between the bosons and $\rho_0$ is the equilibrium 
superfluid number density. The obstacle and 
any applied flow impose appropriate boundary conditions on the condensate
wave function $\psi$.

It will prove convenient to express length in units of the coherence length
$\xi=\sqrt{\hbar^2/g\rho_0 m}$, the condensate in units of $\sqrt{\rho_0}$
and velocities in units of the speed of sound $c=\hbar/m\xi$.

We take as boundary conditions on $\psi$:

\be\label{bc1}
\psi (\vr,t) \rightarrow e^{i v_\infty y}~e^{i \theta}, \qquad r \rightarrow
\infty ,
\ee

\be \label{bc2}
\psi (\vr,t)= 0 , \qquad r \le a .
\ee

The first and second exponentials in \ref{bc1} say that at infinity we
have the external flow along the $y$ direction and the circulation due to a 
vortex somewhere near, or inside the disk.  
We assigned a radius $a$ to the disk and placed the origin at its center.

The determination of the critical velocity for vortex depinning can be
made by looking for steady state solutions of the NLSE equation above.
For $v_\infty$ below a
certain critical value, dependent only on the disk radius, one is able
to find stationary solutions representing a laminar flow. Above the
critical velocity no stationary state exists, a state with a 
circulation around the disk can
evolve to produce a vortex near the edge of the disk. 
The NLSE in $1d$  was found to display the same type of 
behavior \cite{FAL,Hakim}, phase slips are dynamically produced behind 
an inhomogeneity if a large enough flow velocity is applied at infinity. We 
postpone the discussion of the numerical calculation to section \ref{numer}.

\section{The Collective Coordinate Approach}\label{collective}

Now we will follow Ref. \cite{Volovik} and study this problem using as 
a collective coordinate the 
2$d$ location of the vortex in the flow region. We will ignore 
the role of the superfluid compressibility. 

When we write $\psi=\sqrt{\rho}~e^{i\phi}$, the NLSE yields
a continuity equation and an Euler equation for the 
superfluid density $\rho$ and superfluid velocity $\vv=\nabla \phi$. In the
incompressible limit these equations become:

\be \label{e1}  
\nabla \cdot \vv = 0 ,
\ee

\be \label{kelvin}
\frac{\partial \vv}{\partial t} + (\vv \cdot \nabla ) \vv = 0 .
\ee

If a set of vortices of charge $n_i=\pm 1$ is present
at positions $\vR_i$, we also have:

\be \label{e2}
\nabla \times \vv = \sum_i 2\pi n_i \delta (\vr - \vR_i) .
\ee

The boundary condition \ref{bc2} becomes

\be \label{e3}
\vv \cdot \hat{r} = 0 , \qquad r=a .
\ee

The problem of determining the velocity field from a given distribution of
vortices and a fixed boundary condition at infinity is contained in
equations \ref{e1} and \ref{e2}. They can be cast in the form of a $2d$
electrostatic problem if we define the potential $\chi$:

\be
\vv = \nabla \chi  \times \hat{z}.
\ee

Eqs. \ref{e1} and \ref{e2} reduce to a Poisson's equation:

\be\label{poisson}
\nabla^2 \chi = - \sum_i 2\pi n_i \delta (\vr - \vR_i) .
\ee

As for the boundary condition, Eq.
\ref{bc1} translates to $\chi \rightarrow \chi_0 - v_\infty x$ at
infinity and the disk becomes an equipotential of $\chi$ 
due to the impenetrability condition \ref{e3}.

The vortex dynamics is contained in equation \ref{kelvin}. As is well known, 
inside an ideal and incompressible fluid the vortex
follows the flow velocity at its center (excluding the singular 
contribution of the vortex itself to $\vv$). This being simply Kelvin's
theorem \cite{Batchelor}:

\be\label{Ktheorem}
\frac{d\vR}{dt} = \vv (\vR,t).
\ee

In the Appendix we show how to go from Eq. \ref{kelvin} to Eq. \ref{Ktheorem}.

We now derive the dynamics of a vortex of charge $n$ at position $\vR$ 
outside the disk, 
with an applied flow at infinity, $\vv_\infty = v_\infty \hat{y}$, by first
writing the general solution of Eq. \ref{poisson} as the sum of three terms: 

\be\label{chie}
\chi_{\rm e}(\vr ) = -v_\infty x (1-a^2/r^2) ,
\ee

\be
G_0(\vr,\vR ) = -n \ln |\vr-\vR | ,
\ee

\be\label{GI}
G_I(\vr,\vR ) = + n \ln |\vr - (a^2/R^2)\vR | - n \ln (r/a) .
\ee

The total flow, which obeys the topological constraint 
$\oint \vv \cdot d\vl = n$ for a path  enclosing the disk at infinity, 
is given by:

\be
\vv(\vr , t) = \nabla_\vr \left [ \chie (\vr) + G_0 (\vr,\vR ) +
G_I (\vr,\vR ) \right ]  \times \hat{z} .
\ee

To find the flow velocity at the vortex position and the vortex equation
of motion we use the symmetry properties of the Green's functions
and obtain (excluding the singular contribution of $G_0$):

\be\label{eqmotion}
\frac{d\vR }{dt} = \nabla_\vR \left [ \chi_{\rm e}(\vR) + \frac{1}{2} 
G_I (\vR,\vR) \right ] \times \hat{z} .
\ee

One can view this equation
as Newton's equation for a massless object moving in a constant $B$-field along
the $z$ direction plus an electrostatic potential set by $\chi_{\rm e}$ and
$G_I$ (the applied flow plus the vortex images). Alternatively, 
the $X$ and $Y$ coordinates of the vortex can be thought of as conjugate
variables 
with $\chi_{\rm e}+ \frac{1}{2}G_I$ playing the role of the Hamiltonian.

The absence of a mass term is due to the incompressibility assumption,
the effect of the compressibility is to introduce retardation effects in 
the response of the vortex to its external surroundings, which in turn
gives rise to a frequency dependent mass term in the vortex equation
of motion \cite{AF}. The issue of the vortex mass in a neutral superfluid
was, until recently, the subject of many papers \cite{mass}.

Fig. \ref{fig1} shows the isolines of the Hamiltonian function, 
which correspond to the
trajectories followed by a positive vortex in the incompressible limit. 
There one sees a saddle point to the right of the disk
separating closed trajectories near the disk from open ones, of the
same energy, further from it. 

As it stands, Fig. \ref{fig1} describes the dynamics of a 
vortex that is already
outside the disk, but we want to describe the process of vortex depinning
whereby the quantum of circulation, initially inside the disk, 
is pushed outwards by the flow in the form of a vortex. 
The problem we face is characteristic of the collective 
coordinate approach since we are using as variable the very
object whose creation we want to describe. One way out of this dilemma is
to recognize that the incompressibility hypothesis should hold only at
a distance $\delta (\sim \xi)$ away from the obstacle. 
The nucleation process in fact happens, due to compressibility effects, 
within a coherence length from the disk, see Ref. \cite{Frisch} and Fig. 
\ref{fig3} below. 

If the parameter $v_\infty$ is such that the saddle
point is inside this region dominated by the compressibility, one has
the possibility of the nucleated vortex to appear in one of the open
orbits to the right of the disk and be dragged along by the flow, as one 
observes in the simulation of the NLSE. 
Correspondingly,
when the saddle point is outside the compressibility region any nucleated
vortex would remain close to the disk, in one of the closed orbits. We would
like to identify this situation with what happens when the velocity is
below the critical velocity and the quantum of circulation remains 
trapped inside the disk.

We therefore find an estimate for the critical velocity from:

\be
\frac{\partial}{\partial X} \left [ \chie (\vR) + \frac{1}{2} G_I (\vR,\vR ) 
\right ]_{(X,Y)=(a+\delta,0)} = 0 ,
\ee
and view $\delta$ as an adjustable parameter of order $\xi$.

We obtain:
\be\label{simple}
v_{\rm c} = \frac{a^2 (a+\delta)}{(a+\delta)^4 - a^4}.
\ee

An alternative expression for the critical velocity 
is that used in Ref. 
\cite{Josserand}. One writes the NLSE as two equations 
for $\rho$ and $\phi$.
One of the equations is a continuity equation, the other is:

\be
-\frac{\partial \phi}{\partial t} = - \frac{1}{2} \frac{\nabla^2
\sqrt{\rho}}{\sqrt{\rho}} + \frac{1}{2} (\nabla \phi)^2 + \rho -1 .
\ee

In the steady state, $\partial \phi/\partial t$ must be a constant. The
asymptotic region fixes this value and one finds:

\be 
\frac{1}{2} (v_\infty^2 - v^2) = - \frac{1}{2} \frac{\nabla^2
\sqrt{\rho}}{\sqrt{\rho}} + \rho -1 .
\ee

If one ignores the Laplacian in this equation one obtains a relation between
$v$ and $\rho$. Above a certain value of $v$ one has $d[\rho(v)v]/dv < 0$,
this defines $\vmax=\sqrt{(v_\infty^2+2)/3}$ as the maximum possible 
velocity inside the fluid. If at some point $v > \vmax$ a vortex
will be nucleated so as to reduce the flow velocity at that point.
We can estimate the highest velocity in the flow using
the incompressible limit. If a quantum of circulation is present inside
the disk, $v$ is maximum at the equator, and it is equal to
$2 v_\infty + 1/a$. This gives the critical velocity as the maximum value
of $v_\infty$ that ensures $v < \vmax$ everywhere in the flow:

\be\label{pomeau}
\vc=\frac{3}{11}\left [ -\frac{2}{a} + \sqrt{\frac{1}{3a^2}+\frac{22}{9}}
~\right ] .
\ee
 
In the limit $a\rightarrow \infty$ one finds the critical velocity
for vortex/antivortex pair nucleation, $(\vc)_{\rm pair}=\sqrt{2/11}$,
quoted in Ref. \cite{Frisch}.

In Fig. \ref{fig2} we show the dependence of $\vc$ on the disk radius from Eqs.
\ref{simple} and \ref{pomeau}. In the first case we used $\delta=0.5$.

\section{Numerical Simulation}\label{numer}

To determine numerically the critical velocity for depinning we looked
for the values of $v_\infty$ which allowed a steady state solution 
of Eq. \ref{NLSE}. This solution, with the boundary
condition of Eq. \ref{bc1} at infinity, has $e^{-iv_\infty^2 t/2}$ 
as time dependence.
The spatial dependence is then found as a solution of

\be \label{NLSE2}
\frac{v_\infty^2}{2} \psi = -\frac{1}{2}\nabla^2 \psi + (|\psi|^2 -1)\psi,
\ee
with boundary conditions given by Eqs. \ref{bc1} and \ref{bc2}.

We wrote this equation using finite differences on a two dimensional grid 
and reduced the problem to a set of $N$ complex, nonlinear algebraic equations,
for $N$ complex variables, the values of $\psi$ at the grid nodes.
This was solved, starting from a trial state, using Newton's method 
\cite{NumRec}.
The number of nodes varied depending on the disk radius so that we
always had a minimum distance of $10\xi$ between the disk boundary and
the sides of the working area. The grid spacing used was 0.5$\xi$, but
it was reduced to 0.25$\xi$ for disks of smaller
radii to ensure a reasonable representation of the disk as a region
of the square grid. The disk radii ranged from 0.5$\xi$ to $10\xi$.

After a solution
with a given value of $v_\infty$ was found, a new $v_\infty$, increased by
0.1, was used in the boundary condition and in equation \ref{NLSE2}. The
code was run again taking the available solution as trial state.
When Newton's method failed to converge we took the last value of $v_\infty$
as a lower limit for the critical velocity.

The critical velocities so obtained are shown in Fig. \ref{fig2}. 
There one sees
an increase of $\vc$ with the disk radius, as both Eqs. \ref{simple}
and \ref{pomeau} predict, and an apparent saturation at the value 0.5$c$,
higher than the prediction of $\sqrt{2/11}~c$ of Eq. \ref{pomeau}.
The value of $\delta$ that must be used in Eq. \ref{simple} to obtain
this asymptotic value of $\vc$ is $0.5\xi$.

Above the critical velocity, if one uses as initial state a condensate
with a quantum of circulation inside the disk, one can observe the
creation of a vortex from the dynamics of the NLSE, Eq. \ref{NLSE}.
This equation was solved numerically using finite differences in a 
grid similar to the one described above. The condensate at time zero
was evolved using a split-step algorithm \cite{NumRec} where the Laplacian 
part and the nonlinear part of the NLSE were used alternately. The
time step was 0.1$\xi/c$. The condensate amplitude at
four different times is shown in Fig. \ref{fig3}. There one observes the 
vortex formation in the region close to the disk equator and its 
subsequent movement downstream. After the vortex depinning, a vortex/antivortex
pair is seen nucleating as in Ref. \cite{Frisch}. This happens
because the value of $v_\infty=0.5 c$ used is also above the critical
velocity for pair nucleation for that particular radius, $a=1 \xi$.

\section{Macroscopic Quantum Tunneling}

Below the critical velocity an effective barrier
prevents the circulation around the disk to become a vortex  
outside the disk. At zero 
temperature this barrier can only be overcome via Macroscopic Quantum
Tunneling. In a thorough semiclassical
treatment we would have to find the bounce in imaginary time that connects
the condensate corresponding to the stationary, laminar flow, solution of
the NLSE at $\tau=-\infty$ to a condensate containing a nucleated 
vortex near the disk at
$\tau=0$, and back to the stationary state at $\tau=+\infty$. This bounce
would be a nontrivial solution of the imaginary time NLSE \cite{FAL}:

\begin{eqnarray}\label{bounce}
-\frac{\partial \psi}{\partial \tau} &=& - \frac{1}{2} \nabla^2 \psi
+ (\psi\psib - 1)\psi , \qquad \psi(\vr,-\infty)=\psi_0(\vr) , \\
\frac{\partial \psib}{\partial \tau} &=& - \frac{1}{2} \nabla^2 \psib
+  (\psi\psib - 1)\psib, \qquad \psib(\vr,+\infty)=\psi_0^*(\vr) 
\label{bounce2}.
\end{eqnarray}

Here, $\psi_0(\vr)$ is the stationary condensate that represents the
laminar flow around the disk with the circulation trapped inside. The bounce
solution would show, as imaginary time progresses, a vortex leaving
the disk towards the flow and then returning back. The vortex position
at $\tau=0$ would indicate the most favorable position inside the flow 
for the vortex to emerge. 

Instead of retaining all degrees of freedom of the condensate in the 
calculation of the bounce one can use 
the collective coordinate approach of section \ref{collective} 
and focus on the dynamics of the vortex position.
In this approach the bounce, or tunneling path,
connects a closed orbit in the $XY$ plane to its open counterpart having the
same value of $E=\chie + \frac{1}{2} G_I$. 
This is done by allowing the momentum $Y$ to become
imaginary, see Fig. \ref{fig4}. 
The logarithm of the WKB tunneling rate is then proportional to the
$p\dot{q}$ part of the action that governs the vortex dynamics, 
Eq. \ref{eqmotion}. This action in real time is:

\be
S[\vR] = h \rho_0 \xi^2 \int 
\left [ Y \dot{X}  - \chie(\vR) - \frac{1}{2} 
G_I(\vR,\vR)  \right ] ~dt .
\ee

The semiclassical tunneling rate is then:

\be\label{wkb}
\Gamma_{\rm WKB} \propto \exp \left [ - 2 \pi \rho_0 \xi^2 
\oint |Y(X,E)|~dX \right ] ,
\ee
with $Y(X,E)$ being the trajectory in the $XY$ plane that is a solution
(with imaginary $Y$) of:

\be\label{trajec}
\chie(\vR) + \frac{1}{2} G_I(\vR,\vR) = E.
\ee

The argument in the exponential is $2\pi$ times the number of bosons 
enclosed by the imaginary
momentum trajectory. To compare, in the field theory for
the condensate we find the tunneling rate from:

\be
\Gamma_{\rm WKB} \propto \exp \left [- \rho_0 \xi^2 
\int \bar{\psi_{\rm b}}~\frac{\partial \psi_{\rm b}}
{\partial \tau}~~d\vr d\tau \right ],
\ee
where $\psi_{\rm b}(\vr,\tau)$ is the bounce solution of Eqs. \ref{bounce}
and \ref{bounce2}. 

Whereas the semiclassical recipe for the tunneling of $\psi$ is well defined,
in the collective coordinate treatment it is not clear between which 
pair of trajectories the tunneling will take place, or equivalently, which
value of $E$ is to be used in Eq. \ref{wkb}. 
This ambiguity stems from the fact that the collective
coordinate approach does not really describe the vortex nucleation process, 
instead
it describes the tunneling of a preexisting vortex bound to the disk in a
closed trajectory towards an open trajectory further inside the flow.

One can imagine doing the same kind of approximation used to obtain
the formula for the critical velocity and simply admit
that the tunneling will occur between the closed trajectory that reaches 
a distance $\delta$ from the disk and its open counterpart. This
would correspond to the vortex being produced at the furthest possible 
distance inside the compressibility
dominated region around the disk. This 
fixes the value of $E$ to be used in Eq. \ref{wkb} and is consistent with
the formula for the critical velocity, Eq. \ref{simple}, since we then
obtain $\Gamma_{\rm WKB} \rightarrow 1$ as $v_\infty \rightarrow \vc$ (when the
saddle point is exactly at a distance $\delta$ from the disk).

In Fig. \ref{fig5} we show the dependence of the tunneling rate 
with the applied 
velocity, $v_\infty$,
calculated using Eq. \ref{wkb} for radii $a=$ 1, 2 and $3\xi$ and
$\delta=$ 0.5 and 1.5$\xi$.  
One observes, for small values of $v_\infty/v_c$, that

\be
\Gamma \sim \exp \left [ -C~(v_c/v_\infty)^{\alpha} \right ], \qquad 
\alpha=2.8 \pm 0.1 ,
\ee
is a good fit to all curves.

\section{Conclusion}

We have analyzed the process of vortex depinning from a two dimensional 
obstacle due to an applied superflow. We discussed the
process of dynamical depinning, above the critical velocity, and the
process of depinning via MQT, at zero temperature and below the critical
velocity, with a collective coordinate approach. The problems inherent to this
approach were circumvented with the introduction of a length $\delta$ ($\sim
\xi$) proportional to
the size of the compressibility dominated region around the
defect. A more thorough calculation
of the critical velocity was made using the full NLSE 
and compared with the estimate based on the collective coordinate. 
The process of dynamical nucleation was demonstrated 
with a numerical simulation of the time dependent NLSE. 

It is left for a future work to contrast the 
the semiclassical tunneling rate computed from Eq. \ref{wkb} 
with the one obtained by solving the imaginary time NLSE.

Experiments on critical velocities of two dimensional neutral superflows past
obstacles do not exist in the literature, nevertheless one
can expect the calculations presented here to be applicable to layered
superconductors in the superclean limit, when the pancake vortices are
expected to follow the dynamics discussed in section III, see Ref. 
\cite{Blatter}. When such systems are irradiated by heavy ions
defects similar to the ones discussed here can be formed. 
Measurements of the critical current dependence
with the temperature and applied magnetic field already exists, see
for instance Ref. \cite{Prost},
but the dependence with the defect radius is still lacking, when 
obtained a comparison could be made with the results presented here.

\appendix
\section*{Kelvin's Theorem}

Here we derive Kelvin's theorem, Eq. \ref{Ktheorem}, from Eq. \ref{kelvin}.
We consider a flow field $\vv(\vr,t)$ and compute the rate of change
of the circulation in a path $C(t)$ being dragged by the flow:

\be
\frac{d}{dt} \oint_{C(t)} \vv(\vr,t) \cdot d\vl .
\ee

We must evaluate,

\be
\frac{1}{dt} \left [ \oint_{C(t+dt)} \vv(\vr,t+dt)\cdot d\vl -
\oint_{C(t)} \vv(\vr,t)\cdot d\vl \right ] .
\ee

The mapping between points in the $C(t)$ loop and points in 
the $C(t+dt)$ loop is

\be
\vr \rightarrow \vr+\vv(\vr,t)~dt.
\ee
Therefore an element $d\vl$ in $C(t)$ is mapped to
\begin{eqnarray}
\vr + d\vl &\rightarrow & \vr + d\vl + \vv(\vr+d\vl,t)~dt, \\
d\vl &\rightarrow & d\vl + (d\vl\cdot\nabla) \vv(\vr,t)~dt.
\end{eqnarray}

The integral in the $C(t+dt)$ loop can then be written as

\be
\oint_{C(t)}
\vv [ \vr+\vv(\vr,t)~dt ,t+dt] \cdot \{ d\vl + (d\vl\cdot\nabla)
\vv(\vr,t)~dt \} ,
\ee
or, to lowest order in $dt$,
\be
\oint_{C(t)} \{ \vv + \partial_t \vv~dt + (dt~\vv\cdot\nabla)\vv \}
\cdot d\vl + \oint_{C(t)} \vv\cdot(d\vl\cdot\nabla)\vv~dt.
\ee

Finally we obtain for the derivative,
\be
\frac{d}{dt} \oint_{C(t)} \vv\cdot d\vl = \oint_{C(t)} \partial_t \vv
\cdot d\vl + \oint_{C(t)} (\vv\cdot\nabla) \vv \cdot d\vl +
\oint_{C(t)} \frac{1}{2} \nabla v^2 \cdot d\vl .
\ee

The first two integrals add up to zero by virtue of Eq. \ref{kelvin},
whereas the third integral is identically zero as long as the path
does not cross any singularity of $v^2$.

The fact that the circulation is a constant along any path flowing
with the fluid imply that the singular vortices must follow the
flow velocity at their center. That is, Eq. \ref{Ktheorem} must hold.

\begin{figure}
\caption{Isolines of $E=\chie + \frac{1}{2} G_I$, see Eqs. \ref{chie} and
\ref{GI}. These lines also represent 
the trajectories of a positive vortex in the presence of
an applied flow with $v_\infty = 0.4c$ along the $y$ direction. The disk radius
is $a=1\xi$. The solid lines are below the saddle point, the dashed lines
are above.}
\label{fig1}
\end{figure}

\begin{figure}
\caption{Critical velocity, in units of the speed of sound $c$, as a function
of the disk radius $a$. The solid line was obtained from Eq. \ref{simple} with
$\delta=0.5\xi$ and the dashed line from Eq. \ref{pomeau}. The squares are
lower limits for $v_c$ obtained from the time independent NLSE, Eq. 
\ref{NLSE2}, with a grid spacing of $0.5\xi$. The open circles used a grid
spacing of $0.25\xi$ (the two values coincide for $a=0.5\xi$).}
\label{fig2}
\end{figure}

\begin{figure}
\caption{Countour lines of the condensate amplitude obtained from a numerical
integration of the time dependent NLSE, Eq. \ref{NLSE}. The amplitude is
zero inside the disk of radius $2.5\xi$ and increases to the equilibrium
value of $\sqrt{\rho_0}$ away from it. The initial state had a smooth
circulation around the disk in addition to the applied flow in the 
$y$ (vertical) direction of $v_\infty= 0.5c$. The times are 14, 23, 29 and
34$\xi/c$ respectively for figures a), b), c) and d). In a) the
vortex is beginning to form, in b) it is fully formed and being dragged 
by the flow, in c) one observes a vortex/antivortex pair (the antivortex to the
left and the vortex to the right) beginning
to form and fully formed in d).}
\label{fig3}
\end{figure}

\begin{figure}
\caption{Close up view of the saddle point of Fig. \ref{fig1}. Here the
solid lines are the solutions of Eq. \ref{trajec} with real $Y$ and the
dashed lines are solutions with imaginary $Y$. These can also be viewed as
the real and imaginary time trajectories of a positive vortex around
the disk of radius $a=1\xi$ when an applied flow of $v_\infty=0.4c$ along
the $y$ direction is present. The area enclosed by the dashed paths is
proportional to the logarithm of the tunneling rate.} 
\label{fig4}
\end{figure}

\begin{figure}
\caption{Logarithm of the semiclassical tunneling rate as a function of
the applied flow velocity at infinity, calculated with Eq. \ref{wkb}. The
value of $E$ used in that equation was chosen to describe a vortex 
tunneling from a distance $\delta$ from the disk towards
an open trajectory, of same energy, further from it. $v_c$ was
calculated using this value of $\delta$ in Eq. \ref{simple}. 
The squares, circles
and triangles correspond to disk radius $a=$ 1, 2 and 3$\xi$ respectively.
The solid symbols use $\delta=0.5\xi$ and the open symbols use 
$\delta=1.5\xi$.} 
\label{fig5}
\end{figure}

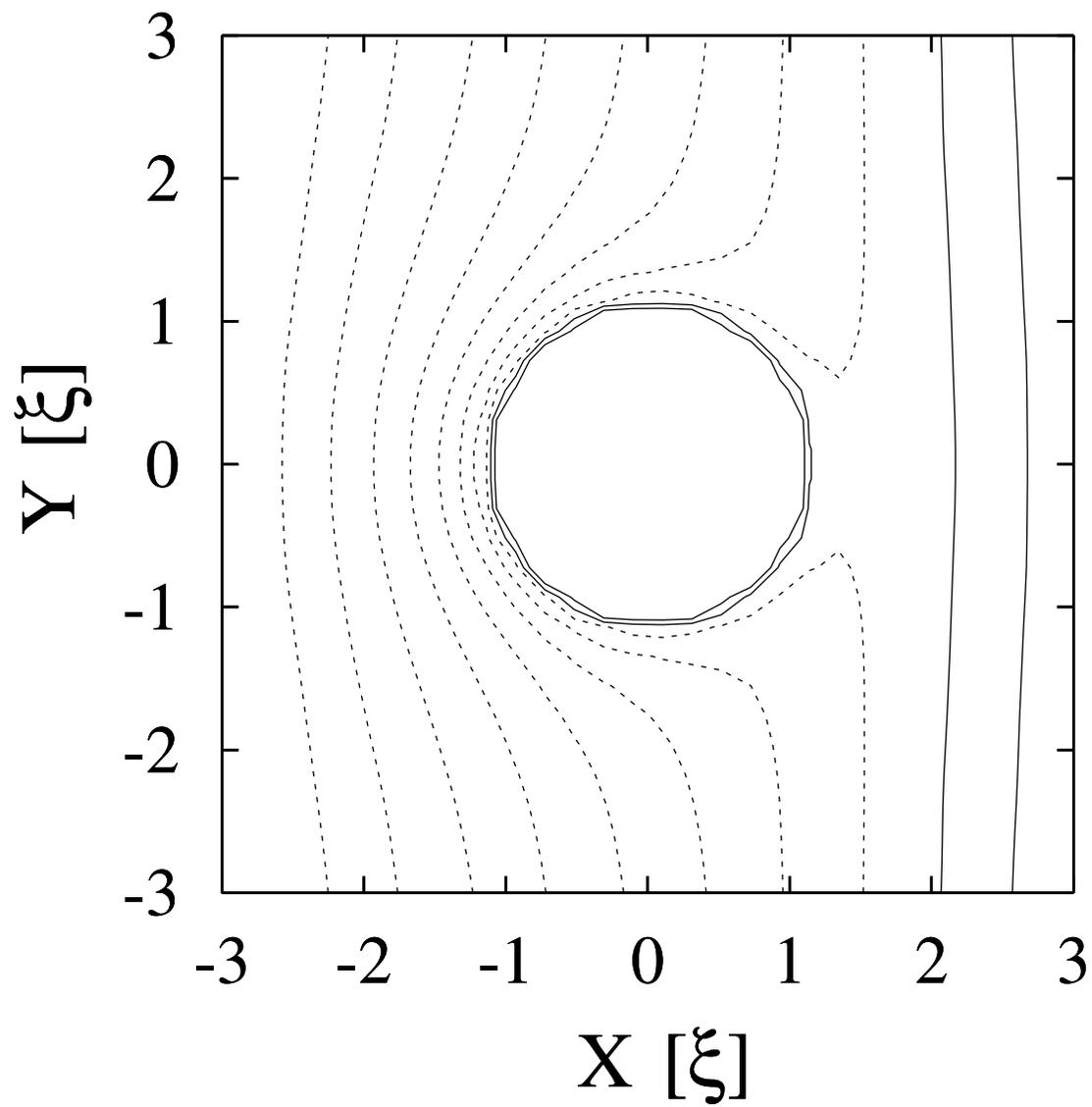
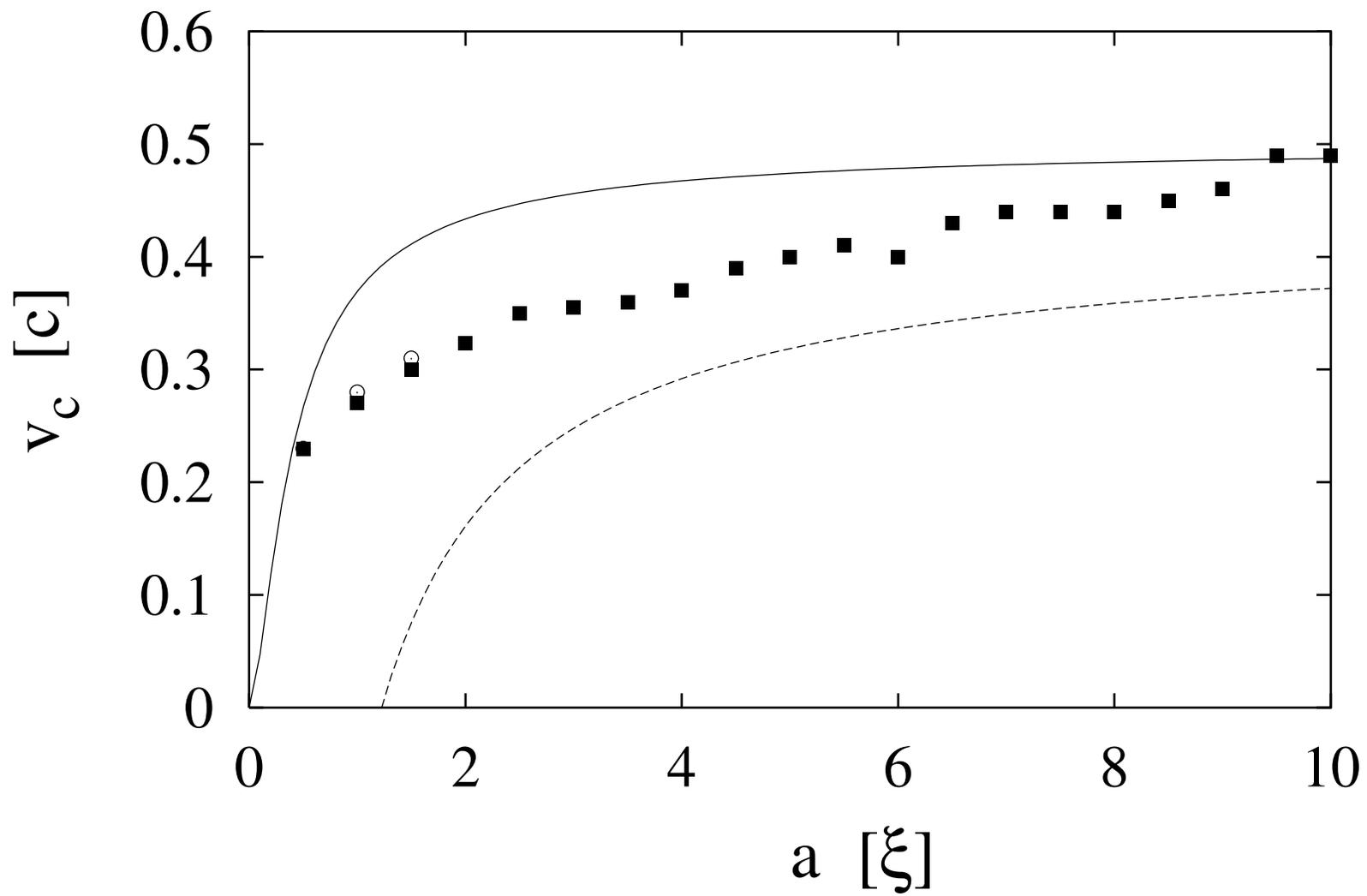
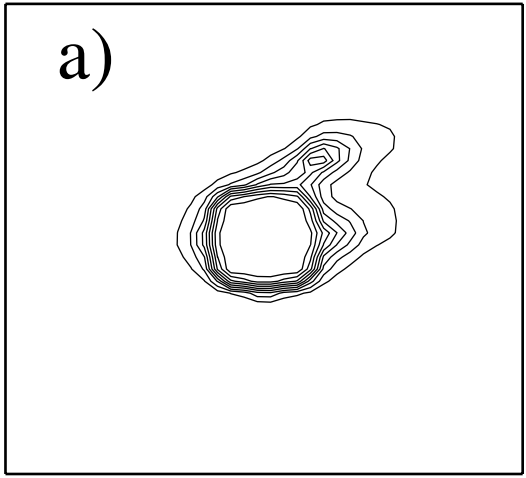
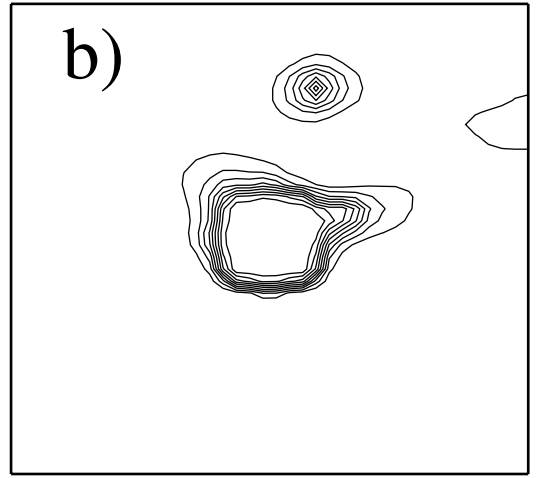
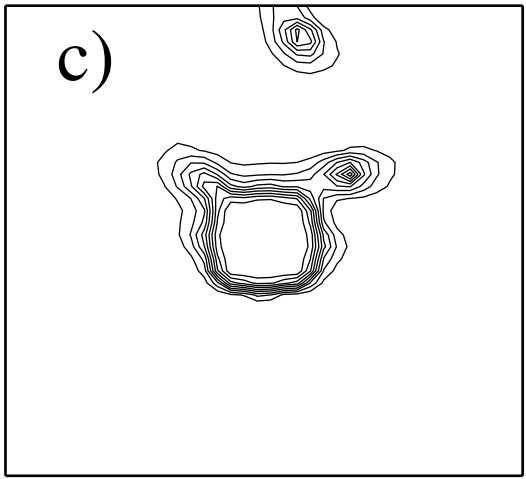
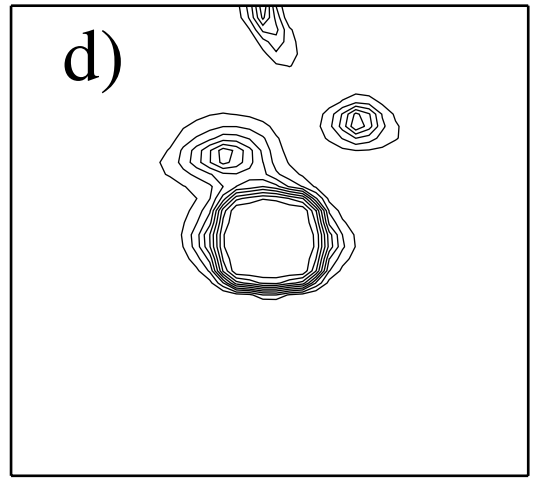
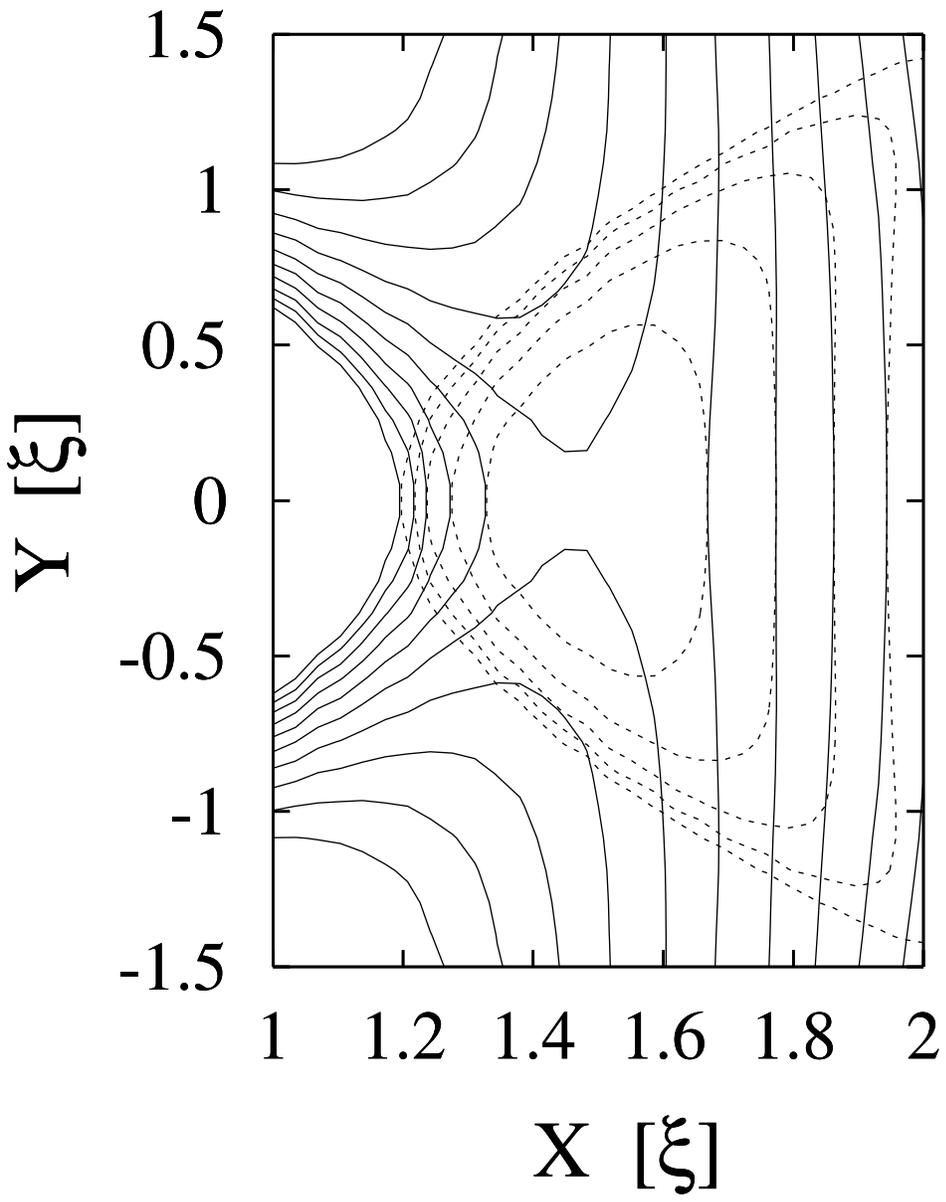
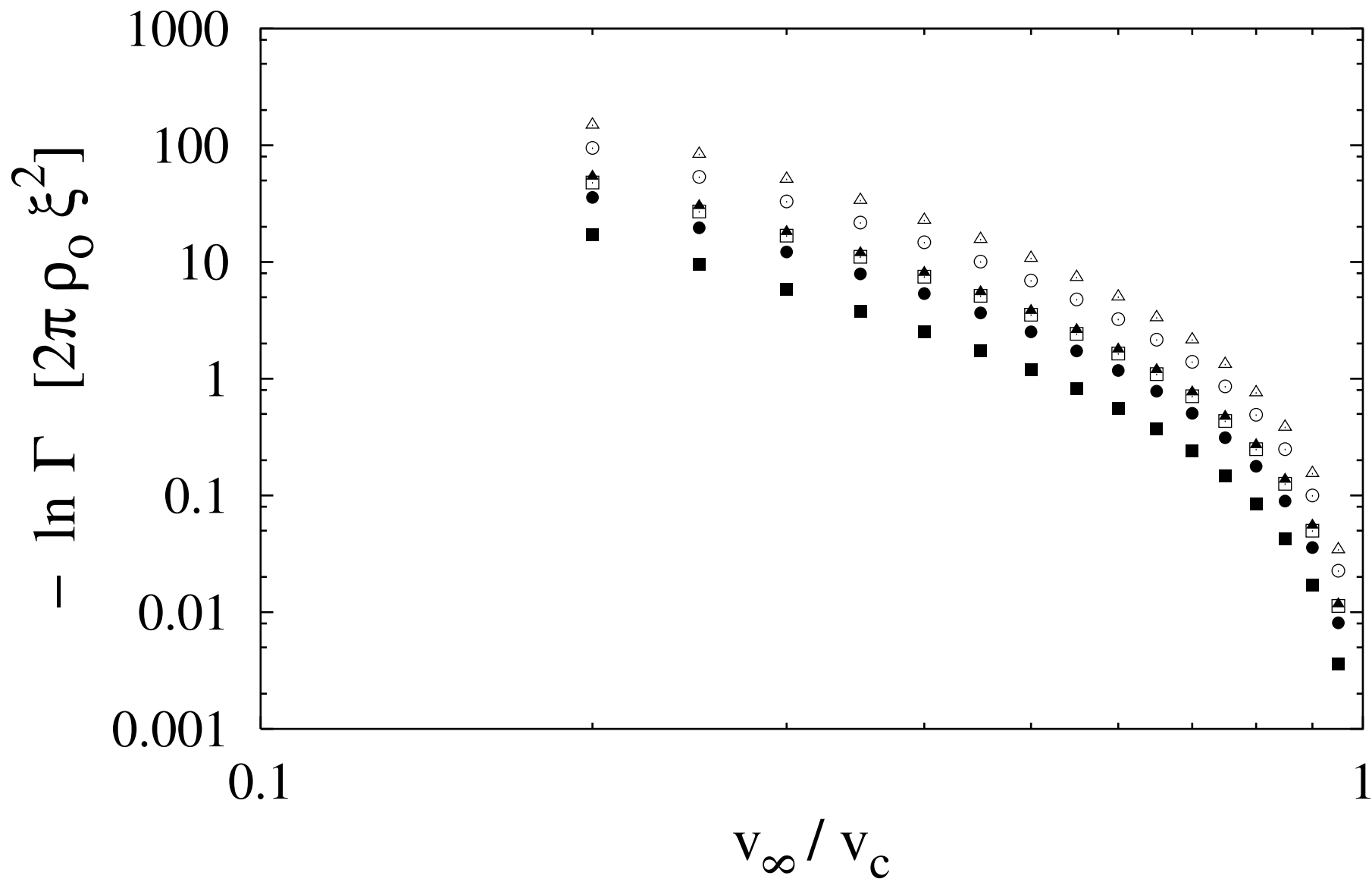
\end{document}